\documentstyle[aas2pp4]{article}

\slugcomment{Submitted to ApJ}

\lefthead{}
\righthead{}

\newcommand{\beq}{\begin{equation}}
\newcommand{\eeq}{\end{equation}}
\def\lsim{\mathrel{\mathpalette\@versim<}}
\def\gsim{\mathrel{\mathpalette\@versim>}}

\begin{document}

\title{On the Radial Structure of Radiatively Inefficient Accretion
Flows with Convection}

\author{Marek A. Abramowicz\altaffilmark{1}, Igor V.
Igumenshchev\altaffilmark{2,3}, Eliot Quataert\altaffilmark{4} and
Ramesh Narayan} \affil{Harvard-Smithsonian Center for Astrophysics, 60
Garden Street, Cambridge MA 02138}

\altaffiltext{1}{Department of Astronomy \& Astrophysics, G\"oteborg
University \& Chalmers University of Technology, 412-96 G\"oteborg,
Sweden}

\altaffiltext{2}{Laboratory for Laser Energetics, University of
Rochester, 250 East River Road, Rochester NY 14623-1299.} 

\altaffiltext{3}{Institute of Astronomy,
48 Pyatnitskaya Ulitsa, 109017 Moscow, Russia.}

\altaffiltext{4}{Institute for Advanced Study, Princeton NJ 08540; Chandra Fellow}

\begin{abstract}

We consider the radial structure of radiatively inefficient
hydrodynamic accretion flows around black holes.  We show that
low-viscosity flows consist of two zones: an outer
convection-dominated zone and an inner advection-dominated zone.  The
transition between these two zones occurs at $\sim 50$ Schwarzschild
radii.

\end{abstract}

%\keywords{globular clusters,peanut clusters,bosons,bozos}

\section{Introduction}

Observations of low luminosity black holes in galactic nuclei and
X-ray binaries have been modeled within the framework of a radiatively
inefficient accretion flow (RIAF).  In these models, radiative losses
are small because of the low particle density of the accreting plasma
at low mass accretion rates (see Narayan, Mahadevan \& Quataert 1998
and Kato, Fukue \& Mineshige 1998 for reviews).

Hydrodynamical simulations of low-viscosity RIAFs in two dimensions
(Igumenshchev \& Abramowicz 1999, 2000; Stone, Pringle \& Begelman
1999) and three dimensions (Igumenshchev, Abramowicz \& Narayan 2000)
have shown that these flows are convectively unstable (see also
Narayan \& Yi 1994, 1995), and that the convection has a significant
effect on the structure of the flow.  Narayan, Igumenshchev \&
Abramowicz (2000, hereafter NIA) and Quataert \& Gruzinov (2000,
hereafter QG) constructed self-similar solutions for
convection-dominated accretion flows (CDAFs), and used these to
explore the relevant physics and to explain the results of the
numerical simulations.

In this paper we show via analytical arguments that the radial
structure of a low-viscosity RIAF consists of two distinct zones: an
outer convection-dominated zone and an inner advection-dominated zone
(\S2).  We present global solutions of an idealized height-integrated
set of equations (\S3) and show that the radial structure of the
advection-dominated zone differs significantly from the canonical
self-similar solution for an advection-dominated accretion flow (ADAF,
Narayan \& Yi 1994, 1995).  This is because of the strong influence of
the boundary conditions.  We then compare the theoretical
results with numerical simulations (\S4) and find that there is good
agreement.  In particular, both the global solutions and the numerical
simulations show that the transition from the inner ADAF zone to the
outer CDAF zone occurs at a radius $R_{AC}\sim 50R_g$, where
$R_g=2GM/c^2$ is the gravitational radius of the accreting black hole
of mass $M$.  
%{\bf (Can we think of another symbol for the transition
%radius?  We have already used $R_{tr}$ for the thin disk-ADAF
%transition. Igor: no problem, but $R_{tr}$ is also OK.)}

\section{Analysis of Self-Similar CDAFs}

We consider a rotating accretion flow with a self-similar scaling of
the height-averaged angular velocity $\Omega$ and isothermal sound
speed $c_s$, and the scale height $H$:
\[
\Omega(R)=\Omega_0 \Omega_K\propto R^{-3/2},
\]
\begin{equation}
c_s^2(R)=c_0^2 v_K^2\propto R^{-1},
\end{equation}
\[
H(R)=c_s/\Omega_K=c_0 R,
\]
where $\Omega_K=\sqrt{GM/R^3}\equiv v_K/R$ is the Keplerian angular
frequency at radius $R$, and $\Omega_0$ and $c_0$ are dimensionless
constants.  We consider a power-law radial dependence for the height
averaged density,
\begin{equation}
\rho(R)\propto R^{-a},
\end{equation}
which corresponds to a radial velocity varying as
\begin{equation}
v_R (R)=\dot{M}/4\pi RH\rho\propto R^{a-2},
\end{equation}
where $\dot M$ is the mass accretion rate.  The index $a$ is equal to
1/2 for a self-similar CDAF (NIA) and 3/2 for a self-similar ADAF
(Gilham 1981; Begelman \& Meier 1982; Narayan \& Yi 1994, 1995).  We
refer the reader to NIA for the calculation of the constants
$\Omega_0$ and $c_0$.

Here we focus on the consistency of the self-similar CDAF
solution. For this purpose we consider the internal
energy equation describing
a RIAF,
\begin{equation}
\rho v_R T{ds\over dR}+{1\over R^2}{d\over dR}(R^2 F_c)=Q^{+}-Q_{rad} \label{energydiv},
\end{equation}
where $s$ is the specific entropy, $F_c$ is the outwardly directed
convective energy flux, $Q^+$ is the rate of energy dissipation per
unit volume, and $Q_{rad}$ is the rate of energy loss per unit volume
through radiation.  By definition, radiative losses are small in a
RIAF, so we neglect $Q_{rad}$ in equation (4).  The first term on the
left hand side of equation (4) describes the inward advection of the
internal energy.  For the scaling laws given in (1) and (2) this
advection term takes the form,
\begin{equation}
\rho v_R T{ds\over dR}={\dot{M}c^2\over 4\pi RH}{c_0^2\over 2}
{R_g\over R^2}\left({1\over\gamma -1}-a\right),
\end{equation}
where $c$ is the speed of light and $\gamma$ is the adiabatic index of
the accreting gas.  For a rotating accretion flow, the energy
dissipation rate $Q^+$ can be written as
\begin{equation}
Q^+=-(J_v+J_c){d\Omega\over dR},
\end{equation}
where $J_v$ and $J_c$ are the angular momentum fluxes due to viscosity
and convective motions, respectively. In the case of an ADAF
($a=3/2$), convection is ignored and we set $J_c=F_c=0$.  Since
viscosity transports angular momentum outwards, i.e. $J_v>0$, the
dissipation rate $Q^+$ is a positive quantity, and in equation (4),
$Q^+$ balances the advection term (5).

The situation is very different in a CDAF.  NIA considered a
self-similar CDAF in the limit of a non-accreting ``convective
envelope'', in which $v_R=0$ and the advection term (5) vanishes.  How
can a viscous differentially-rotating fluid configuration not accrete?
This is possible because, as demonstrated explicitly via
three-dimensional numerical simulations (Igumenshchev et al. 2000) and
earlier through analytic methods (Ryu \& Goodman 1992; Stone \& Balbus
1996), axisymmetric
convection transports angular momentum inwards, i.e. $J_c<0$,
whereas viscosity transports it outwards.  Thus, by ensuring that
convection and viscosity cancel each other exactly, i.e.  $J_v+J_c=0$,
the convective envelope is able to achieve a configuration with no net
angular momentum transport and hence no mass accretion.  Since
$J_v+J_c=0$, an additional consequence is that $Q^+=0$ (see eq.~[6]).
The energy equation (4) then has only one term left, namely the
divergence of the convective flux, and the only way to satisfy this
equation is for the divergence of the flux to vanish, i.e. to have
$F_c(R)\propto R^{-2}$.

The non-zero outward energy flux $F_c$ corresponds to a ``convective
luminosity'', $L_c=4\pi RH F_c$, which is independent of radius. The
source of this luminosity is not specified in the self-similar
solution and is formally located at $R=0$.  NIA argued that in the
case of a real CDAF there would be a small but non-zero accretion
rate, $\dot{M}\ne 0$, and the convective luminosity would be supplied
by the gravitational release of binding energy in the innermost region
of the accretion flow.  Thus, we may write
\begin{equation}
L_c=\varepsilon\dot{M}c^2, \label{eff}
\end{equation}
where $\varepsilon$ is the ``convective efficiency''.  A non-zero
accretion rate requires a non-zero net angular momentum flux directed
outward, $J_v+J_c>0$, so according to (6) one would also have 
$Q^+>0$.\footnote{Taking into account that the release of
gravitational and rotational energy
is the ultimate source of any dissipation in the accretion flow,
one can conclude $Q^+\propto (GM/R)\rho v_R/R\propto R^{-4}$.}

%{\bf DO WE NEED THIS PARAGRAPH?? (ELIOT) I AGREE THAT WE MIGHT WANT TO
%DELETE THIS PARA (RAMESH). Igor: Equation (8) is shown here only to
%demonstrate the dependence $Q^+\sim R^{-4}$. This needs to discuss an
%importance of different terms in eq.(4). We can skip this paragraph if
%one finds another way to estimate $Q^+$.  ELIOT: I am inclined to
%delete and assume the reader is happy with the scaling or put this in
%a footnote in the appropriate place where the scaling is used in the
%next paragraph} One can rewrite equation (6) for the energy
%dissipation in another form, as the fraction $\eta$ ($0<\eta<1$) of
%the gravitational energy released at the given radius $R$,
%\begin{equation}
%Q^+=\eta\rho {v\over R} {GM\over R}=
%{\dot{M}c^2\over 4\pi RH}{\eta\over 2}{R_g\over R^2}.
%\end{equation}
%The value of $\eta$ depends on the structure of the accretion flow.
%For example, in the case of Shakura-Sunyaev solution for thin accretion
%disks one has $\eta=1/6$.

%{\bf (Igor: I have changed this paragraph.)}  
We now estimate the
importance of different terms in the energy equation (4).  For this we
integrate equation (4) over a spherical volume from $R$ to infinity.
After the integration, the terms which correspond to the advection
term (5) and the dissipation term (6) both scale $\propto R^{-1}$,
whereas the term describing the convective energy transport is equal
to $L_c=$ constant.  Thus, convective energy transport dominates the
physics at large radii, and the self-similar CDAF solution is a good
approximation for describing this region of a RIAF.  At small radii,
however, the advection term and the viscous dissipation term become
more important.  By comparing the inward advective energy flux,
\begin{equation}
\dot{E}_{adv}=\int_{R}^{\infty}4\pi RH\,\rho v_R T{ds\over dR}\, dR=
{c_0^2\over 2}\left({1\over\gamma -1}-a\right){R_g\over R}\dot{M}c^2,
\end{equation}
with the convective luminosity $L_c$ (eq.~[7]), we may determine the
transition radius $R_{AC}$ that separates the outer
convection-dominated zone and the inner advection-dominated zone.
Using the estimate of $c_0^2$ from QG or NIA's Appendix, and applying
the condition $\dot{E}_{adv}=L_c$, we obtain
\begin{equation}
{R_{AC}\over R_g}\sim{1\over\varepsilon}{3 - \gamma\over 2(\gamma+3)}.
\end{equation}
For a quantitative estimate of $R_{AC}$ we need to determine the value
of the convective efficiency parameter $\varepsilon$.  Since it is
hard to determine this from first principles, we must either solve for
$\varepsilon$ as an eigenvalue of the global problem (\S3) or
determine it from numerical simulations (\S4).
%From (10) one has $R_{tr}=(0.24/\varepsilon)R_g$ for $\gamma=5/3$ and
%$R_{tr}=(0.51/\varepsilon)R_g$ for $\gamma=4/3$.

\section{Global CDAFs}

%The self-similar CDAF solution described in \S2 formally has zero
%radial velocity, i.e., $v_R = 0$.  This is incompatible with the
%requirement of black hole accretion that the gas flow supersonically
%into the black hole. 
In this section we present a sample height integrated global CDAF
model which satisfies proper inner and outer boundary conditions.
This is the CDAF analogue of the global ADAF models calculated by
Narayan, Kato, \& Honma (1997) and Chen, Abramowicz, \& Lasota (1997).
The solution sheds further light on the CDAF-ADAF transition discussed
in \S2.

In the absence of radiation, the conservation of mass, angular
momentum, and total energy can be expressed via conserved fluxes: \beq
\dot M = - 4 \pi R H \rho v_R = {\rm const}, \label{mass} \eeq \beq 4
\pi F_L = -\dot M R^2 \Omega + 4 \pi RH(J_v + J_c) = {\rm const}, \eeq
\label{ang} \beq 4 \pi F_E = -\dot M B + 4 \pi R H \Omega (J_v +
J_c) + 4 \pi R H F_c = {\rm const}, \label{energy} \eeq where $B = 0.5
v_R^2 + \phi + 0.5 R^2 \Omega^2 + \gamma c^2_s/(\gamma - 1)$ is the
Bernoulli function, and we take the Paczy\'nski \& Wiita (1980) form
of the gravitational potential, $\phi = -GM/(R-R_g)$, to mimic a
Schwarzschild black hole.  Note that the radial derivative of equation
(\ref{energy}) yields equation (\ref{energydiv}) with $Q_{rad} = 0$.
In addition to equations (\ref{mass})-(\ref{energy}), we need the
radial momentum equation: \beq v_R{dv_R \over dR} = R\Omega^2 - {d
\phi \over dR} - {1 \over \rho} {dp \over dR},
\label{radial} \eeq where $p = \rho c^2_s$ is the pressure.

We use the expressions for $J_v$, $J_c$ and $F_c$ from NIA's Appendix
except that we employ a causal prescription for the convective
transport of energy and angular momentum.\footnote{For numerical
reasons, we were unable to find global solutions without employing a
causal transport prescription.  This is somewhat unsatisfying
because the causal prescription explicitly imposes an inner ADAF zone
on the global CDAF solution.}  We take the convective diffusion
coefficient to be $\propto (1-v_R^2/c_c^2)^2$ for $|v_R| < c_c$ and
$0$ otherwise (cf. Popham \& Narayan 1992; Narayan 1992), where we
choose $c_c = \alpha c_s$ (i.e. roughly the convective velocity).  The
condition $|v_R| = c_c$ defines a transition radius, $R_{AC}$, inside
of which there is no convective transport of angular momentum and
energy.  For $R < R_{AC}$, the inflow velocity of the gas exceeds the
characteristic convective velocity, and convective transport is less
important than bulk advective transport (the convective transport
vanishes in our simplified model).  Outside of $R_{AC}$ the inflow
velocity of the gas is less than the characteristic convective
velocity and the flow is roughly ``static'', as in the self-similar
CDAF solution.

We note that $J_v$ depends on the value of the Shakura-Sunyaev
viscosity parameter $\alpha$, so the global solution depends on the
choice we make for this parameter.  Other parameters are the adiabatic
index $\gamma$ and the outer radius of the accretion flow $R_{out}$.
In addition, although the causal prescription for convective
transport is well motivated (e.g., Narayan 1992), the proper cutoff
velocity (i.e., $c_c$) is somewhat uncertain.

Equations (\ref{mass})-({\ref{radial}) are four first order
differential equations.  Since we are interested in solutions in which
the gas goes through a sonic point on its way into the black hole, the
sonic radius $R_s$ is an additional unknown which needs to be
determined. The 3 conserved fluxes, $\dot M$, $F_L$, and $F_E$ are
also eigenvalues determined in a self-consistent manner.  This implies
that the convective efficiency $\varepsilon$ of equation
(\ref{eff}) is an eigenvalue of the global model, rather than a free
parameter as in the self-similar solution.  The transition radius
$R_{AC}$ is a final eigenvalue.  For $R < R_{AC}$, $F_c = 0$ and $J_c
= 0$ because of the causal prescription for convective transport.  One
can show that the vanishing of $F_c$ means that $F_E$ and $F_L$ are
not independent, $F_L = F_E (R^2 \Omega/B)|_{R = R_s}$}.  Thus,
we require a total of 8 boundary conditions to solve for the 4
variables ($\rho$, $v_R$, $c_s$, $\Omega$) and 4 independent
eigenvalues ($R_s$, $R_{AC}$, $\dot M$, $F_E$).

At the outer radius of the flow we specify $\rho$, $c_s$, and
$\Omega$.  The choice of $\rho$ is arbitrary and simply sets the
scaling for the density in the problem.  We use the self-similar CDAF
values for $c_s$ and $\Omega$ from QG at the outer boundary.
Igumenshchev \& Abramowicz (1999, 2000) showed from numerical
simulations of RIAFs that strong convection is present only for
relatively small values of the viscosity parameter, $\alpha<0.1$.  For
such values of $\alpha$, $R_{AC} > R_s$.  At the sonic point $R=R_s$,
convection is thus unimportant and the 2 boundary conditions at this
radius are identical to those in the global ADAF models of Narayan et
al. (1997).  We also apply the no-torque condition at the sonic
point.\footnote{Narayan et al. (1997) applied the no-torque condition close
to the black hole horizon; we do not expect this to make a large
difference.}  The final two boundary conditions are at the transition
radius $R_{AC}$.  We guess a value for $R_{AC}$ and find the solution
from the outer boundary to $R_{AC}$ and from $R_{AC}$ to the inner
boundary (the sonic point), requiring $|v_R| = c_c$ at $R_{AC}$.  We
then iterate until all flow variables are continuous across $R_{AC}$.

Figure 1 shows the density, radial velocity, and convective velocity
$c_c$ for a global solution with $\alpha = 0.03$ and $\gamma = 1.5$.
The sonic point of the solution is at $R \approx 2.6 R_g$, while the
transition radius is located at $R_{AC} \approx 35 R_g$.  The angular
momentum flux in the solution is $4 \pi F_L \approx - 1.64 \dot M R_g
c$ while the energy flux is $4 \pi F_E \approx 0.0045 \dot M c^2$.  At
large radii the energy flux is entirely carried by convection, so this
corresponds to a convective efficiency of $\varepsilon \approx 0.0045
$.  For this value of $\varepsilon$ and $\gamma=1.5$, equation (9)
predicts $R_{AC}=37R_g$, which is very close to the numerically
determined location of the transition radius.

\begin{figure}
\plotone{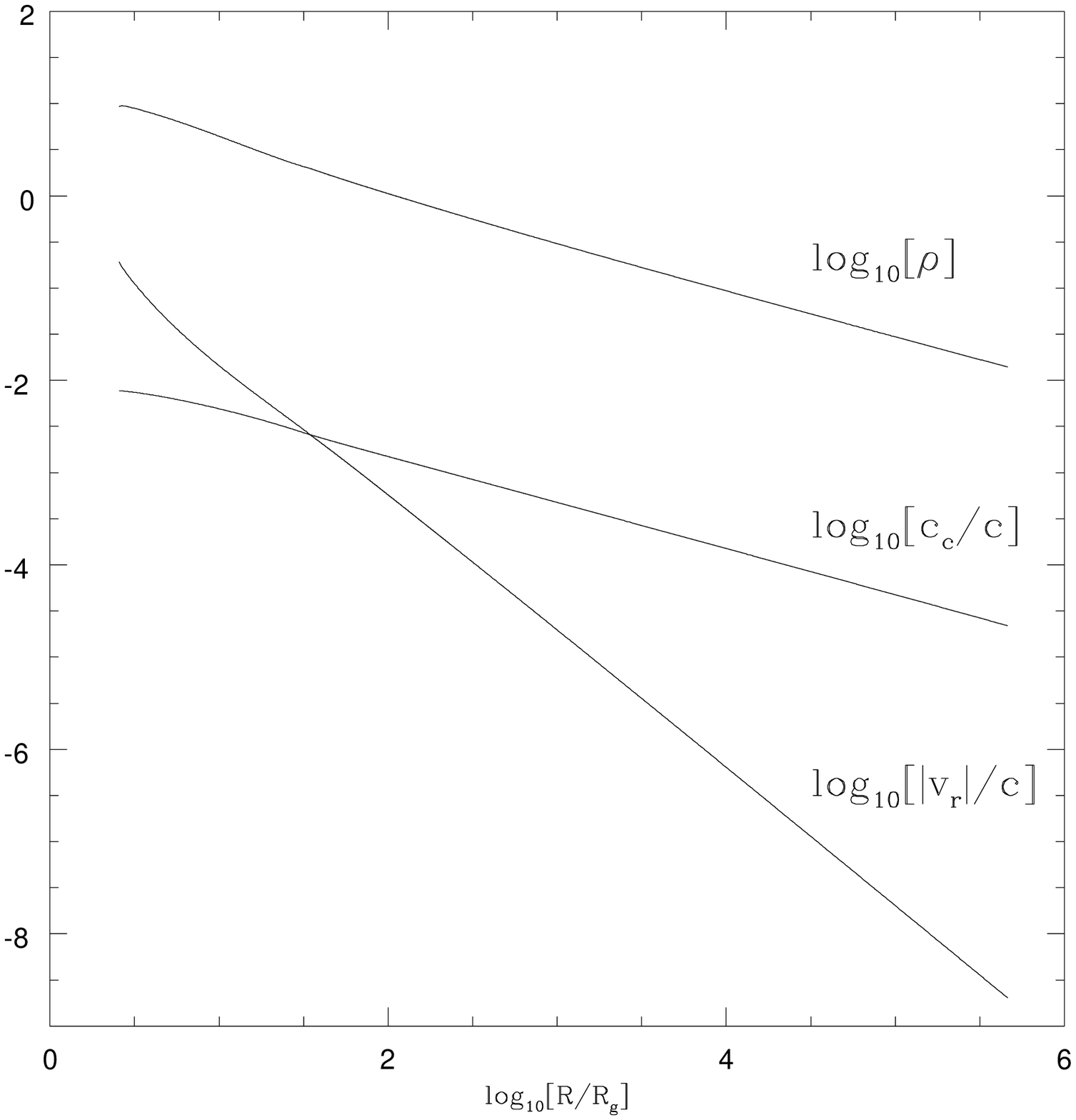}
\caption{Global CDAF model with $\alpha = 0.03$ and $\gamma = 1.5$.
The density profile $\rho$ is in arbitrary units.
Inside $R = R_{AC} \approx 35 R_g$ the inflow speed exceeds the local
convective velocity ($c_c$) and energy transport in the flow is
dominated by inward advection.  At larger radii, the flow is roughly
static ($|v_r| \ll c_c$) and energy
%transport is dominated by convection. \label{fig1}}
is transported outward by convection. \label{fig1}}
\end{figure}

\placefigure{fig1}

The density profile in Figure 1 is reasonably well described by the
$\rho \propto R^{-1/2}$ power law of the self-similar CDAF solution at
all radii.  For $R > R_{AC}$ convective transport indeed dominates the
energy transport and the scaling $\rho \propto R^{-1/2}$ is needed in
order to carry a constant convective luminosity outwards (see \S2 and
NIA, QG); in this regime the self-similar CDAF solution describes the
flow structure quite well.

For $R < R_{AC}$, however, convective transport of energy and angular
momentum vanishes in our causal prescription.  Here, the radial inflow
velocity exceeds the convective velocity and bulk (advective)
transport of energy dominates.  The density profile is nonetheless
close to the CDAF scaling of $R^{-1/2}$ and much flatter than the
$R^{-3/2}$ profile of a self-similar ADAF.  The reason for this is
that the gas is rapidly accelerating toward the sonic point on its
way into the black hole.  Quite generally, this rapid acceleration
leads to a steep radial velocity profile and thus a flat density
profile (see, e.g., Narayan et al.'s 1997 and Chen et al.'s 1997
global ADAF models for similar results).

\section{Comparison with Numerical Simulations}

Igumenshchev (2000) and Igumenshchev \& Abramowicz (2000) found that
the convective efficiency is $\varepsilon\simeq 0.003-0.01$ in
two-dimensional simulations of RIAFs, with only a weak dependence on
the viscosity parameter $\alpha$ and the adiabatic index $\gamma$.  We
expect that three-dimensional models of convective RIAFs will show
values of $\varepsilon$ in the same range, because of the close
qualitative and quantitative similarities between the numerical
results in two and three dimensions (Igumenshchev et al. 2000).
Taking a characteristic value of $\varepsilon=0.005$, we obtain from
equation (9): $R_{AC}\simeq 30 \, R_g$ for $\gamma=5/3$ and
$R_{AC}\simeq 40 \, R_g$ for $\gamma=4/3$.  These values for
$\epsilon$ and $R_{AC}$ are in good agreement with the values obtained
with the global model described in \S3.  We now try to estimate
$R_{AC}$ directly from the numerical simulations.

The presence of two distinct zones in a RIAF, namely an outer
convection-dominated zone and an inner advection-dominated zone, may
be demonstrated in a numerical simulation by considering the ``mass
inflow rate'' $\dot{M}_{in}$ as a function of radius.  
%Following Stone et al. (1999), 
We define $\dot{M}_{in}(R)=\int\rho v_R\,dS$, where the
surface integral is limited to those gas elements at radius $R$ that have
$v_R<0$.  Because a CDAF is convectively turbulent and accretes
slowly, roughly half the mass at any radius at any time will be
flowing in and half will be flowing out. This gives
$\dot{M}_{in}\simeq 4\pi RH\rho\tilde{v}/2$, where $\tilde{v}$ is the
rms velocity of the turbulent eddies. It can be shown that
$\tilde{v}\sim \alpha v_K \propto R^{-1/2}$ (NIA).  Therefore, for
$\rho\propto R^{-1/2}$ and $H\propto R$, one finds that
$\dot{M}_{in}\propto R$.  In an ADAF on the other hand, convection is
absent and so the mass inflow rate is independent of radius,
$\dot{M}_{in}(R)={\rm const}$.

Figure~2 shows $\dot{M}_{in}(R)$ (solid line) for a two-dimensional
simulation of a RIAF with $\alpha=0.01$ and $\gamma=5/3$ (the
numerical technique is similar to that used by Igumenshchev \&
Abramowicz 2000).  This simulation was done with the Paczy\`nski \&
Wiita (1980) pseudo-potential.  The plot clearly shows that the
accretion flow consists of two different zones: an inner
advection-dominated zone, in which $\dot{M}_{in}={\rm const}$, and an
outer convection-dominated zone, in which $\dot{M}_{in}\propto R$. The
transition radius between the two zones is located at $R_{AC}\simeq 50
\, R_g$, which is roughly consistent with the scaling given in
equation (9).

\begin{figure}
\plotone{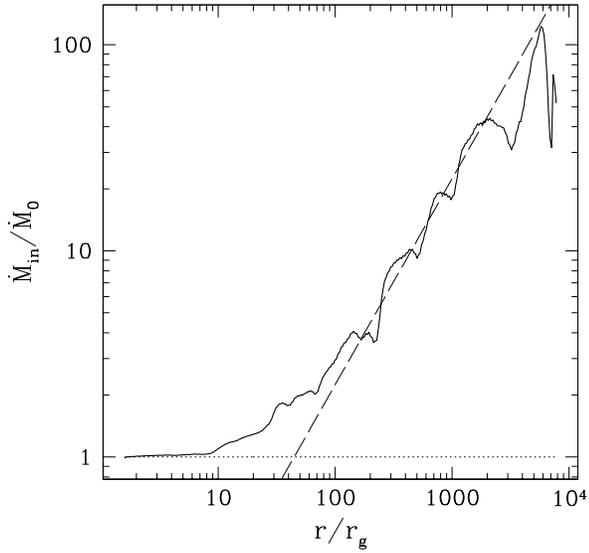}
\caption{Instantaneous mass inflow rate $\dot{M}_{in}$ (solid line)
as a function of radius $R$ in a 2D hydrodynamical model of a
radiatively inefficient accretion flow with $\gamma=5/3$ and
$\alpha=0.01$. The values of $\dot{M}_{in}$ are calculated by adding
all the inflowing gas elements at a given $R$.  $\dot{M}_{in}$ is
normalized to the net accretion rate $\dot{M}_{0}$.  Except near the
outer boundary, $R \ga 2\times 10^3 R_g$ and in the inner region,
$R\la 10^2 R_g$, the profile $\dot{M}_{in}(R)$ shows good agreement
with the prediction of the self-similar solution for
convection-dominated accretion flows, $\dot{M}_{in}\propto R$, shown
by the long-dashed line.
%In the inner region, $R\la 10^2 R_g$, the flow is 
%advection-dominated, in which $\dot{M}_{in}\simeq{\rm const}$.
Deviation of $\dot{M}_{in}$ from the self-similar CDAF scaling 
in the inner region is due to the strong inward advection
of internal energy, as discussed in the text.
\label{fig2}}
\end{figure}

\placefigure{fig2}

Figure~3 shows a snapshot of the velocity streamlines of the
particular numerical simulation analyzed in Fig. 2.  We see numerous
vortices associated with the convective motions in the flow.
Interestingly, unlike in the oversimplified global model described in
\S3, we see that convective motions are present even in the inner
regions of the RIAF ($R < R_{AC}$), despite the dominance of advection
here.
%Unless advection dominates in the innermost part of RIAFs, there still
%convective motions are present. 
In fact, the convective blobs primarily originate in the
advection-dominated part and move outward.  According to our estimate
(see equation [5] and [7]), the major energy release happens in the
advection-dominated part of RIAFs, where the advection and dissipation
terms are almost equally important and dominate the convective energy
transport.  Only a small relative excess of the energy dissipation
rate over the inward advection of energy supports turbulence in RIAFs
and provides the convective luminosity $L_c$.  This excess determines
the value of the convective efficiency $\varepsilon$.

\begin{figure}
\plotone{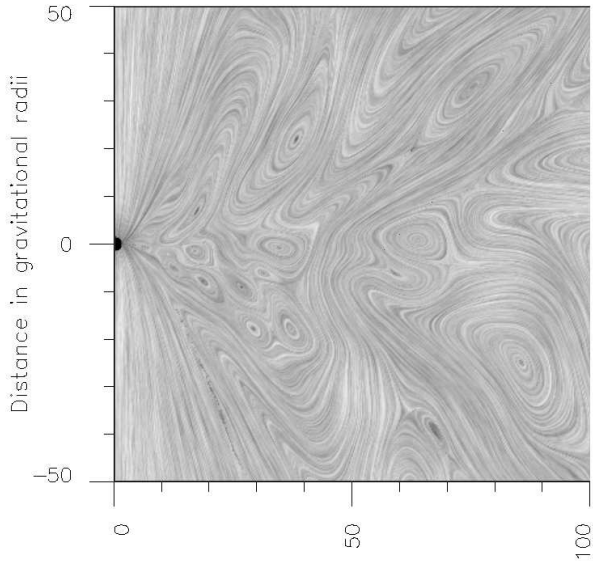}
\caption{Snapshot of streamlines from a 2D hydrodynamical model of a
radiatively inefficient accretion flow with $\alpha=0.01$ and
$\gamma=5/3$.  A projection of streamlines on the meridional cross-section 
is shown, with the black hole
located at the origin. The flow pattern is highly time dependent
and consists of numerous temporal vortices of different spatial scales. In the
innermost region of the flow the energy balance is dominated by inward
advection of internal energy.  Convective blobs originate in this
region and move toward the outer convection-dominated region at radii
$\ga 10^2 R_g$.
\label{fig3}}
\end{figure}

\placefigure{fig3}

%  Include ????
%  
%The self-similar ADAF solution is scale-free, whereas the boundaries
%introduce two characteristic radial scales: the inner scale $\simeq
%2-3 \,R_g$, inside which the accretion matter is almost in free-fall,
%and the outer scale $R_{tr}\sim 10^2 R_g$.  It seems that the radial
%range between these two scales is too small for the appearance of 
%self-similar behavior for the advection-dominated solution.

\section{Discussion and Conclusions}

%{\bf Eliot: I modified this somewhat} 
We have considered the radial
structure of radiatively inefficient accretion flows (RIAFs). Based on
an analysis of self-similar solutions we conclude that the flow
consist of two parts: an outer convection-dominated part where the
radial velocity is highly subsonic and an inner advection-dominated
part where the gas flows rapidly inwards.  The location of the
transition radius $R_{AC}$ between these two parts depends primarily
on the convective efficiency $\varepsilon$ (see equation [9]).  We
estimate $\epsilon \approx 0.003-0.01$ and $R_{AC} \sim 50 R_g$
directly from numerical simulations (\S4).  In addition, in 1D steady
state global calculations of the structure of RIAFs, which include
proper boundary conditions, $\epsilon$ and $R_{AC}$ can be determined
as eigenvalues.  The global calculations yield $R_{AC} \approx 35 R_g$
and $\epsilon \approx 0.0045$ (\S3), in good agreement with the
simulations.

Convective motions and the associated outward energy flux originate in
the inner advection-dominated part of the flow.  This energy flux
supports the turbulent structure of the flow in the outer
convection-dominated part.  The radial profiles of density and
velocity in the inner advection-dominated part of the flow differ
significantly from self-similar scalings due to the influence of
boundary conditions (\S3).  

%{\bf IGOR: HEREAFTER IT HAS BEEN MODIFIED.}  
Our present results are
based on a viscous hydrodynamical (HD) approach to studying RIAFs.
The HD approach is not self-consistent because it requires one to
assume the presence of viscosity with an {\it a priori} unknown
strength.  A fully consistent approach can be provided within the
framework of magneto-hydrodynamics (MHD), in which torques due to
magnetic interactions transport angular momentum.  It is not presently
clear how the concept of CDAFs obtained in the HD approach will be
modified after construction of MHD models of RIAFs.  

We do not expect significant qualitative differences between HD and
MHD models if the gravitational and rotational energies are released
locally in the accretion flow, e.g., by magnetic reconnection and/or a
turbulent energy cascade. If, however, the free energy is removed from
the bulk of the accretion flow via non-local interactions, such as the
stretching of large-scale magnetic field lines resulting in an
efficient Poynting flux, or the formation of MHD driven
winds/outflows, one would expect qualitative differences between HD
and MHD models.  In addition, one would expect a significant
modification of the CDAF solution if angular momentum is efficiently
removed from the accretion flow by a large-scale magnetic field in a
disk corona (contrary to the CDAF model in which angular momentum
transport is due to local viscous interactions).  

MHD simulations of spherical accretion flows (Igumenshchev \& Narayan
2001) support the idea of local generation of energy via magnetic
reconnection, which leads to the development of efficient
convection. The structure of these models closely resembles CDAFs.
Recent MHD simulations of rotating flows do show unstable motions,
which leads to the development of efficient turbulence, but they do
not show clear behaviour to either support or contradict the CDAF
model.  For example, Machida, Matsumoto, \& Mineshige (2001) claim
that the structure of their MHD models closely resembles that found in
HD CDAFs. However, Hawley, Balbus, \& Stone (2001) stress that their
models do not demonstrate convective motions and that the models are
instead turbulent due to the magneto-rotational instability in
shearing flows.  This discrepancy indicates that further development
of the MHD models is required.

Recently, Balbus (2001) has discussed the stability criteria for
rotating magnetized accretion flows. These criteria replace the
H{\o}iland criteria for a plasma with thermal conduction along the
magnetic field lines. According to Balbus's criteria a magnetized
fluid is unstable even if it is marginally stable to the H{\o}iland
criteria.  The new criteria have been obtained in the short-wavelength
approximation, and it is not straightforward to know how the
instability works on larger scales, which might be more important for
the structure of an accretion flow. It is also interesting to clarify
what Balbus's criteria imply for axisymmetric convection in the
presence of a predominantly toroidal magnetic field. Such convection
is expected to be present in magnetized CDAFs.

%({\bf Ramesh: I have added some discussion of spectra.  Please change
%or delete as you wish.})  
Assuming that the radial structure of RIAFs
described in the present paper is basically correct, it is interesting
to ask what the implications are for observations of accreting black
holes.  It has been proposed that for luminosities below a few percent
of the Eddington luminosity, accretion flows around black holes switch
to a RIAF (cf. Narayan et al. 1998).  In the scenario proposed by
Narayan (1996) and developed by Esin, McClintock \& Narayan (1997),
the accretion flow consists of a radiatively efficient thin disk
outside a transition radius $R_{tr}$ and an ADAF inside $R_{tr}$.
Spectra computed with such a model agree well with observations of
black hole X-ray binaries in the ``intermediate state'' and ``low
state'' (Esin et al. 1997, 1998).  The models typically require
$R_{tr}$ of order $10R_g$ to a few tens of $R_g$.  In a recent study,
Esin et al. (2001) determined that the black hole binary XTE J1118+480
had $R_{tr}\sim50R_g$ at a time when its X-ray luminosity was
$\sim10^{-3}L_{Edd}$.  According to the present paper, an ADAF is a
good approximation to the structure of a RIAF so long as $R\la50R_g$.
Therefore, the above models are consistent.

For yet lower luminosities, e.g., black holes in the ``quiescent
state'', the transition radius in models is typically at $R_{tr} \sim
10^3-10^4R_g$, and in some models there is no thin disk at all (see
Narayan et al. 1998 for examples).  Published models of the quiescent
state have generally made use of a pure ADAF (or a spherical Bondi
flow, e.g. Melia 1992), whereas our present results suggest that the
flow might take the form of a CDAF for $R\ga50R_g$.  How will this
affect the predicted spectrum?  The principal difference between an
ADAF (or a Bondi flow) and a CDAF is the value of the index $a$ 
(cf eq.~[2]) in the radial density profile: $a=3/2$ for a self-similar ADAF,
$a=1/2$ for a self-similar CDAF.  Quataert \& Narayan (1999) studied
models with different values of $a$ and showed that these models can
produce similar spectra provided the fraction of viscous heat energy
that goes into electrons (which is assumed to be close to zero for an
ADAF) is appropriately adjusted.  For a CDAF, perhaps as much as half
the heat energy will need to go into the electrons to match
observations (e.g. Ball, Narayan \& Quataert 2001).  Detailed models
are awaited.

\acknowledgments This work was supported in part by NSF grant AST
9820686, RFBR grant 00-02-16135, and Swedish NFR grant.  EQ
acknowledges support provided by NASA through Chandra Fellowship grant
number PF9-10008 awarded by the Chandra X-ray Center, which is
operated by the Smithsonian Astrophysical Observatory for NASA under
contract NAS8-39073.

\appendix

\clearpage

% Now comes the reference list.  In this document, we used \cite to call
% out citations, so we must use \bibitem in the reference list, which
% means we use the LaTeX thebibliography environment.  Please note that
% \begin{thebibliography} is followed by a null argument.  If you forget
% this, mayhem ensues, and LaTeX will say "Perhaps a missing item?" when
% you run it.  Do not call us, do not send mail when this happens.  Put
% the silly {} after the \begin{thebibliography}.
%
% Each reference has a \bibitem command to define the citation format
% to be placed in the text (in []) and the symbolic tag used for 
% cross referencing (in {}).
%
% See sample1.tex, or the AASTeX guide, for an alternative to the \cite-
% \bibitem command.

\clearpage

\clearpage

\clearpage

\end{document}